\documentclass[conference]{IEEEtran}
\IEEEoverridecommandlockouts
\usepackage{cite}
\usepackage{amsmath,amssymb,amsfonts}
\usepackage{algorithm}
\usepackage{algorithmic}
\usepackage{graphicx}
\usepackage{textcomp}
\usepackage{xcolor}
\usepackage{xspace}
\usepackage{soul}
\usepackage{booktabs}
\usepackage{subfig}
\usepackage{graphicx}

\def\BibTeX{{\rm B\kern-.05em{\sc i\kern-.025em b}\kern-.08em
    T\kern-.1667em\lower.7ex\hbox{E}\kern-.125emX}}
\begin{document}

\title{Secure Aerial Surveillance using Split Learning}

\author{\IEEEauthorblockN{Yoo Jeong Ha, Minjae Yoo, Soohyun Park, Soyi Jung, and Joongheon Kim}
\IEEEauthorblockA{Department of Electrical and Computer Engineering, Korea University, Seoul, Republic of Korea 
\\
E-mails: 
\texttt{\{mj7015,soohyun828,jungsoyi,joongheon\}@korea.ac.kr}
}
}
\maketitle

\begin{abstract}
Personal monitoring devices such as cyclist helmet cameras to record accidents or dash cams to catch collisions have proliferated, with more companies producing smaller and compact recording gadgets. As these devices are becoming a part of citizens' everyday arsenal, concerns over the residents' privacy are progressing. Therefore, this paper presents SASSL, a secure aerial surveillance drone using split learning to classify whether there is a presence of a fire on the streets. This innovative split learning method transfers CCTV footage captured with a drone to a nearby server to run a deep neural network to detect a fire's presence in real-time without exposing the original data. We devise a scenario where surveillance UAVs roam around the suburb, recording any unnatural behavior. The UAV can process the recordings through its on-mobile deep neural network system or transfer the information to a server. Due to the resource limitations of mobile UAVs, the UAV does not have the capacity to run an entire deep neural network on its own. This is where the split learning method comes in handy. The UAV runs the deep neural network only up to the first hidden layer and sends only the feature map to the cloud server, where the rest of the deep neural network is processed. By ensuring that the learning process is divided between the UAV and the server, the privacy of raw data is secured while the UAV does not overexert its minimal resources.  
\end{abstract}

\section{Introduction}
The dismal admonition from George Orwell's dystopian fictional masterpiece \emph{1984}, "Big Brother is watching you," became the reality people of the twenty-first century live in. Progressively, cities and citizens rely on street-level surveillance to help fight off larceny, arson, hate crimes, or any crimes against persons, property, or public order. 
Surveillance unmanned aerial vehicles (UAVs) are sought after by law enforcement agencies for their ability to watch the area at diverse angles without any blind spots {\cite{uav1,uav2,tvt202106jung,9467353}}. Increased surveillance comes hand in hand with increased data uploaded to the cloud. Insurance companies, construction firms, city councils all need closed-circuit television (CCTV) footage to ensure a safer experience on the streets. Furthermore, CCTV scenes are widely utilized in criminal detection and often revealed to the public through the media. Yet, the majority of these data from CCTV cameras on UAVs include car plate numbers, people in vulnerable situations, and even some ghastly scenes of accidents. 

Overcoming the issue of exposing potentially sensitive information contained in the captured CCTV videos, this paper considers split learning {\cite{9026781,pieee202105park}}. A couple of extra points must be considered in this surveillance UAV scenario as the data source comes from a power-restricted mobile device. Since these CCTV footage are captured by energy limited mobile devices, (i.e., UAVs), and solely interact with a centralized server through wireless data transmissions, visual data uploading are considered {\cite{tmc201907koo,video1, mm2017koo,video2,ton201608kim,video3,jsac201806choi}}. The surveillance UAVs travel around the city with limited resources, hence, UAV caching {\cite{video2, twc201912choi, twc202104choi, tmc202106malik, twc202012choi}} and fast uploading {\cite{8119812,7045578,7961189,tvt2021jung}} of feature maps to the server are important factors to be considered in this split learning algorithm.

\begin{figure*}[h!]
    \begin{center}
        \includegraphics[width=0.95\linewidth]{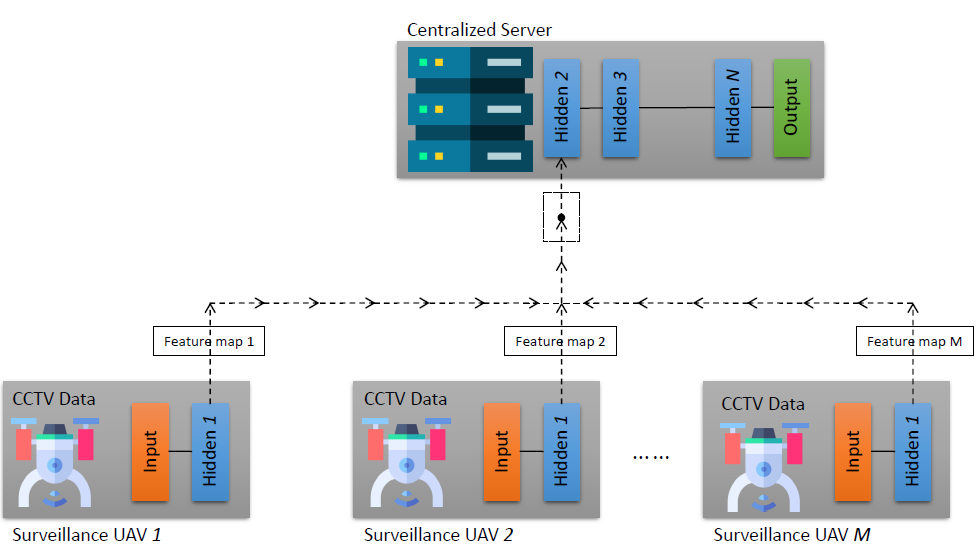}
    \end{center}
    \caption{ The overall architecture of our proposed SASSL model.}
    \label{architecture}
\end{figure*}

This paper places an exceptional emphasis on preserving the privacy of original CCTV footage captured by surveillance UAVs.
Split learning is extremely effective in situations where the original data must be protected at all costs {\cite{pieee202105park}}. The structure of the split learning algorithm ensures the privacy of the original data since only the encrypted feature map is transferred to the server. The surveillance drone runs the deep neural network only up to the first hidden layer, where the raw footage captured from the UAV becomes distorted. The feature map, which is highly deformed information, is uploaded to the server from the first hidden. The server then uses this feature map as the input to the deep neural network. Since the only exposed information is the encrypted feature map, which is considerably distorted, privacy is guaranteed. The framework of our proposed SASSL, \textbf{S}ecure \textbf{A}erial \textbf{S}urveillance using \textbf{S}plit \textbf{L}earning, algorithm is meticulously drawn in Fig. {\ref{architecture}}. Even if malicious threats target the surveillance drone, the only information revealed is the ciphered feature map. It is nearly impossible for hackers to trace back to the original data from the feature map.    


\section{Model}
\subsection{System Model}
The main construction of our proposed SASSL is the tangible split or division of the learning process. As shown in Fig.~{\ref{architecture}}, there is a distinct division into two parts: the surveillance UAV (i.e., end-system) and server. Due to constraints based on the energy consumption of a mobile device, it is more resource-efficient and comprehensive for the UAV to offload data to a nearby server. During this offloading process, the UAV sends the feature map that is extracted after it runs through the first hidden layer, which is placed within its own system. This distorted feature map about the original image is delivered to the server, where it can fully process the deep neural network; depending on the model selected, the server will classify, predict or detect particular objects. By allowing the UAV to run the deep neural network up to the first hidden layer and transfer the feature map to the server, the privacy of the original data is preserved. 

We introduce a scenario where there are three UAV systems, also known as the end-system, and only one centralized server. We selected three to be the number of end-systems since it adequately emulates a realistic scenario where multiple drones are deployed in action to monitor a suburb. As there are multiple UAVs collecting footage of the neighborhood, the problem of data-imbalance and overfitting becomes insubstantial. In other words, if one UAV captures an insufficient amount of fire occurring in the streets, then the model trained using this small quantity of data is bound to be overfitted. However, since this particular UAV is involved in a split learning approach, it acquires the chance to cooperate with other UAVs that possess an abundant amount of data. All of the multiple UAVs connected to the server collaboratively work to build a deep neural network located up in the server.  

\subsection{Network Model}
The above section explored how the system model is literally split between the end-systems and server. The following section will investigate how the deep neural network (i.e., the network) is divided between each end-system and server.

For an efficient power allocation of UAVs, the data contained within the mobile surveillance system is offloaded to a nearby server. Since the portable UAV is not a high-spec device that can run an entire deep neural network on its own while still roving the streets, it runs the neural network only up to the first hidden layer. As depicted in Fig. {\ref{architecture}}, the feature map extracted from the UAV level is uploaded to the server. The feature map collected from each of the three UAVs is then concatenated, and the integrated feature map becomes the input to the server. The server trains the deep neural network.

The essential idea in this paper is to preserve the privacy of the raw footage captured by the airborne surveillance camera. The original footage obtained by the surveillance UAV is contained within its internal system, and only the encrypted feature map is sent to the server, thereby protecting the raw CCTV image that includes sensitive information. 

In order to deliver the data from UAVs to a server, millimeter-wave (mmWave) communications can be used for high-speed low-delay networking~\cite{tvt2021jung,twc201910choi,6955961}.

\section{UAV Split Learning Framework}
Using the proposed SASSL method, we devised a scenario where three surveillance drones roam a suburb to monitor flames. A convolutional neural network (CNN) is selected as the deep neural network. It is used to classify whether or not there is a fire present in the neighborhood. Once the surveillance UAV hovers around the city, the captured CCTV images are sent to the server in the form of a feature map after running through on hidden layer in the internal system.   

The CCTV images of a suburb with and without flames is a dataset obtained from Kaggle\footnote{https://www.kaggle.com/ritupande/fire-detection-from-cctv}. It is a large dataset that contains flame and non-flame images in the city and the wildlife. Out of this curation, we utilized 331 flame images captured from CCTVs and 533 non-flame images captured from CCTVs to train CNN.

With our SASSL configuration, multiple surveillance drones, all with independent energy constraints, can be considered. However, for the purpose of this paper, we devise an environment based on these conditions: 
\begin{itemize}
    \item There are three surveillance UAVs.
    \item All three surveillance UAVs run the CNN model up to the first hidden layer.
    \item The ratio of the data distribution for split learning is set to 7:2:1.
    \item The energy and battery status is assumed to be the same for all three UAVs.
\end{itemize}

The CNN model setting used to classify whether or not the captured CCTV footage from the surveillance UAV contains flames are summarized in Table {\ref{tab:cnn}}.

\begin{table}[t] 
\begin{center}
	\centering
	\begin{tabular}{l|ccc}
    \toprule[1.0pt]
    \centering
       Parameters & CCTV Flame \\
    \midrule[1.0pt]
    Epochs  & 50 \\
     Loss & Binary crossentropy \\
    Activation function & Sigmoid \\
     Batch size & 32  \\
    Input Size & 64 $\times$ 64 $\times$ 1 \\
    Model & Custom CNN with VGG19 \\
    \bottomrule[1.0pt]
	\end{tabular}
\end{center}
\caption{The CNN setup to determine whether the CCTV image contains flames or not.}
\label{tab:cnn}
\end{table}






\section{Performance Evaluation}

\begin{figure}[t]
    \begin{center}
        \includegraphics[width=0.95\linewidth]{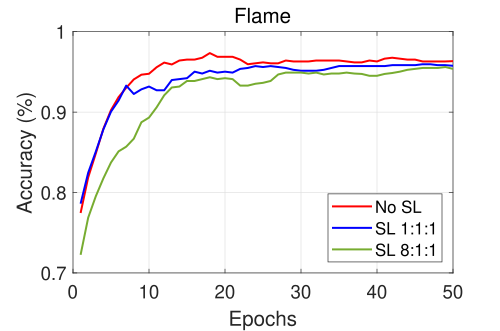}
    \end{center}
    \caption{Experimental results}
    \label{exp}
\end{figure}

The significant advantages of SASSL are, as continuously emphasized throughout this paper, the protection of raw data and effective handling of data-imbalance. This section proves the superiority of our SASSL algorithm compared to the standard deep learning method. 
We experimented with the setup outlined in the above section and obtained Fig.~{\ref{exp}}.

The green curve represents the split learning method with data-imbalance. To demonstrate the efficacy of split learning regarding data-imbalance, we separated the data into an 8:1:1 ratio. In other words, one UAV will possess 80\% of the data, the second UAV will have 10\% of the data, and the last UAV will possess 10\% of the data. The red curve depicts the split learning method without the issue of a data-imbalance considered. Thus, the experiment for the red curve is conducted with the three UAVs holding an equally distributed amount of data. The blue curve portrays the conventional deep learning method where the deep neural network is trained in one fixed position. This curve is obtained by disregarding the data-imbalance case by setting the data ratio to 1:1:1 and running the entire CNN model on the server. Hence, this method completely exposes the original CCTV data that may contain sensitive personal information to the external network; there is no safeguard on the raw footage.

\begin{figure} 
\center
    \subfloat[\centering Original CCTV flame image]{
    \includegraphics[width=0.4\linewidth]{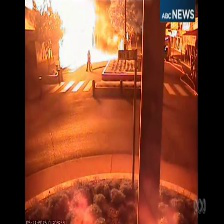}
    }
        \qquad
    \subfloat[\centering Distorted CCTV flame image]{
    \includegraphics[width=0.4\linewidth]{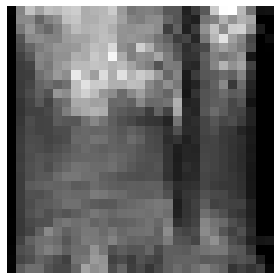}
    }
    \caption{ The (a) original image of a fire captured by a surveillance camera and (b) the feature map that is processed by the first hidden layer at the UAV level.}
    \label{img:feature}
\end{figure}

It should be noted that the infinitesimal drop in classification accuracy that comes with split learning can be neglected given the vast advantages the multi-client split learning brings--privacy preserving and overcoming data-imbalance. The conventional deep learning approach performs the best, achieving the highest classification accuracy of 96.53\%. The split learning method with equally distributed datasets recorded a classification accuracy of 95.95\%. Our proposed SASSL technique, with a data-imbalance of an 8:1:1 ratio, attains a performance level of 95.37\%, which is only a 1.16\% degradation in classifying whether there is a fire outburst in the city compared to the traditional approach. Fig. {\ref{img:feature}} (a) is the original CCTV footage of a fire in the neighborhood. Fig. {\ref{img:feature}} (b) is the distorted feature map that becomes the input to the server. Evidently, the characteristics of the original image are completely lost in (b); thus, preserving any private information contained in the raw CCTV footage. Hence, it shows that a very insignificant decline in performance is sacrificed for a much greater benefit of securing personal information.

\section{Conclusions and Future Work}\label{sec:sec5}
The fleet of surveillance UAVs helps stop illegal acts of drug offenses, burglary, and even monitoring extreme weather conditions. In this paper, we constructed an environment where surveillance UAVs look for fire in a neighborhood. Fire can be initiated as an act of crime, a genuine accident in the home, or a wildlife bushfire. This paper strives to achieve an effective fire classification model without sacrificing sensitive information in these CCTV images. Due to the architectural framework of our proposed SASSL, the raw CCTV footage is protected since only the feature map is exposed to the server. Furthermore, with our SASSL model, the issue of data-imbalance is no longer present. As the deep learning model is trained collaboratively with feature maps from multiple UAVs, even if one UAV has insufficient images of the city, the CNN can still be trained well-using data obtained from other UAVs. SASSL shows a minuscule lower classification accuracy than a complete model. Yet, this comes with an even more significant advantage of privacy preservation.  

Since UAVs are mobile devices, they are limited in resources. Further research can be extended to consider the battery status, power-allocation, and even joint distributed link scheduling of the end-systems. After real-time classification of fire initiation, the network can be set to map the precise location of the fire and immediately call the nearest fire department for swift handling of the hazard before the flames spread.

\section*{Acknowledgment}
This work was supported by the National Research Foundation of Korea (NRF) grant funded by the Korea government (MSIT) (No. 2021R1A4A1030775). J. Kim is a corresponding author of this paper.

\bibliographystyle{IEEEtran}
\bibliography{ref_arvr,ref_aimlab,ref_uav}
\end{document}